\documentclass[12pt]{article}
\usepackage[english]{babel}
\usepackage{array}
\usepackage{float}
\usepackage{booktabs}
\usepackage{amsmath}
\usepackage{amsthm}
\usepackage{amssymb}
\usepackage[a4paper]{geometry}
\usepackage[bookmarks=false,  breaklinks=false,pdfborder={0 0 1},backref=section,colorlinks=false]{hyperref}


\begin{document}
\title{Ultrametricity of Energy Minimum Configurations of RNA Secondary Structures in the Nussinov Model}
\author{A.\,P.~Zubarev \\
 \textit{Volga State Transport University,} \\
 \textit{Pervyi Bezymyannyi pereulok 18, Samara, 443066, Russia;}
\\
 \textit{Samara University,} \\
 \textit{Moskovskoe shosse 34, Samara, 443123 Russia} \\
 e-mail:\thickspace{}\texttt{apzubarev@mail.ru} }
\maketitle
\begin{abstract}
We present a numerical study of the ultrametric properties of the set of RNA secondary structures with the maximum number of base pairs (energetically degenerate minima) within the maximum matching model (Nussinov algorithm). Using 18 natural small nuclear RNAs as examples, we show that the degree of nontrivial ultrametricity varies widely. We consider the optimization problem for the degree of nontrivial ultrametricity of RNA secondary structures with the maximum number of base pairs under a fixed nucleotide composition. It is found that permuting the nucleotide sequence can strongly change the degree of ultrametricity, indicating the key role of nucleotide order in shaping the hierarchical properties of RNA secondary structures.
\end{abstract}

\section{Introduction}

\label{sec_intro}

The concept of ultrametricity, originally arising in functional analysis
and number theory, has acquired fundamental significance in the physics of
complex systems in recent decades. An ultrametric space is characterized by
a strengthened triangle inequality, wherein any three points form an isosceles
triangle with a base not exceeding the legs.
Such a geometry naturally describes hierarchically organized systems
in which elements group into clusters, clusters group into higher-level
clusters, and so on \cite{rammal1986}.

Ultrametricity played a key role in the theory of spin glasses.
Works \cite{parisi1979,parisi1980,mezard1984,mezard1987}
showed that the space of equilibrium states of the Sherrington--Kirkpatrick
model possesses an ultrametric structure. This result (subsequently proven
mathematically rigorously, see, e.g., \cite{talagrand2003,talagrand2011,panchenko2013})
demonstrated that ultrametricity is not an artificial mathematical
construction but a fundamental property of the energy landscape of disordered
systems with many competing minima.
The success of the ultrametric approach in spin glass physics influenced
related fields, especially biophysics. More than forty years ago,
Frauenfelder hypothesized the ultrametric nature of proteins. By analyzing
the results of a series of experiments on carbon monoxide (CO) binding to
myoglobin \cite{frauenfelder1988,frauenfelder1991,frauenfelder1999},
Frauenfelder suggested that a protein can exist in many different
metastable substates organized hierarchically: substates separated by small
energy barriers are grouped into clusters separated by higher energy barriers,
these clusters are grouped into even larger clusters, and so on.
Such a hierarchical structure is naturally described by an ultrametric tree,
and the dynamics of the protein as a random walk on this ultrametric tree.
The study of the hierarchical organization of the energy landscape of
biopolymers naturally led to the application of $p$-adic analysis methods (see \cite{ALL,ALL_1,DKKM} and references therein).
Ultrametric spaces arising in $p$-adic mathematics proved to be an adequate
language for describing the dynamics of complex hierarchical systems. In
works \cite{avetisov2002,ABZ_2014,avetisov2009,bikulov2021}, a connection was
established between the hierarchical structure of the protein energy landscape
and $p$-adic parametrization, and ultrametric diffusion on such landscapes was
investigated. These results showed that $p$-adic models are not a formal
apparatus but reflect the deep hierarchical nature of the organization of
biological macromolecules. In parallel, ultrametric and $p$-adic analysis
methods have also found application in related fields of mathematical biology.
Work \cite{kozyrev2008} constructed a model of the genetic code based on the
representation of codons as states on the plane of $p$-adic numbers. This
direction was developed in \cite{dragovich2009,khrennikov2009,dragovich2012,khrennikov2012,dragovich2021},
which showed that many properties of DNA and RNA, including mutational
stability and evolutionary trajectories, admit a natural interpretation in
terms of $p$-adic ultrametric geometry. Thus, an understanding has now formed
that ultrametric hierarchy is a universal principle manifesting both at the
level of the physical organization of the energy landscapes of biopolymers
and at the level of the structural organization of genetic information.

In 1996, Higgs investigated the ultrametric properties of RNA secondary
structures in his work \cite{higgs1996}. An RNA molecule is a polymer chain
composed of a sequence of four types of nucleotides: adenine (A), cytosine (C),
guanine (G), and uracil (U) \cite{brion1997,draper1999}. Nucleotides are called
complementary if they can form a stable hydrogen bond between the corresponding
nitrogenous bases according to the Watson--Crick base pairing rule: adenine
(A) is complementary to uracil (U), and guanine (G) is complementary to cytosine
(C). The RNA molecule can fold into compact structures by forming hydrogen
bonds between complementary bases \cite{tinoco1971,crothers1974}. Each base can
participate in at most one pair. Pair formation stabilizes the molecule,
lowering its energy \cite{freier1986,mathews1999}. An RNA secondary structure
is a set of base pairs satisfying certain rules \cite{zuker1989,hofacker1994}.
The role of RNA in the cell is extremely diverse. Besides classical functions
(messenger, transfer, and ribosomal RNA), there are numerous functional
non-coding RNAs, such as small nuclear RNAs (snRNAs) involved in pre-mRNA
splicing, small nucleolar RNAs (snoRNAs) guiding ribosomal RNA modifications,
the signal recognition particle (SRP RNA) involved in protein transport
across the endoplasmic reticulum, and RNase P RNA, which possesses catalytic
activity and carries out pre-tRNA processing \cite{golden1998,kiss2001,hainzl2002}.
Each of these functional RNAs must fold into a specific three-dimensional
structure required for its biological activity. The folding energy landscape
of RNA determines which secondary structures are stable and how the molecule
reaches its native conformation.

The present study builds on the work of Higgs \cite{higgs1996}, which
investigated a sample of structures generated with Boltzmann weights at
different temperatures. In \cite{higgs1996}, it was found that at low
temperature (e.g., $T=0.1$), the matrix of pairwise distances between
structures exhibits properties that Higgs interpreted as ultrametric.
Higgs used an ultrametricity criterion consisting of calculating the ratio
$s/s_{\text{unc}}$, where $s$ and $s_{\text{unc}}$ are the average difference
between the largest and middle sides of correlated and uncorrelated triangles,
respectively. This criterion shows that triangles in real data are on average
more isosceles than random ones. Understanding the degree and nature of
ultrametricity of the RNA energy landscape is important for several reasons.
First, ultrametricity can be directly related to the possibility of fast and
directed folding of the RNA molecule. In other words, the hierarchical
organization of the landscape creates a cascade of energy barriers, allowing
the molecule to fold not by random search through all possible conformations
(combinatorial explosion), but through sequential assembly: first, stable local
structures form (overcoming low barriers), and then they assemble into the
global conformation (overcoming higher barriers) \cite{wolynes1995}.
Furthermore, hierarchical organization may reflect evolutionary constraints:
functionally important RNA regions must possess a certain degree of freedom
for conformational rearrangements, which may prevent the achievement of a
completely ultrametric landscape \cite{fontana1993,ancel2000}.
Finally, quantifying ultrametricity can serve as an indicator of the
"smoothness" or "ruggedness" of the landscape, which is important for modeling
folding kinetics.

The aim of this work is to conduct a numerical study of ultrametricity
on the set of strict energy minima (ground states) for a number of reference
RNAs from the NCBI database \cite{refseq}. We consider the classical model of
RNA secondary structure, where each formed base pair lowers the structure's
energy by one unit, so the total energy of structure $\alpha$ is
$E(\alpha)=-N_{\text{pairs}}(\alpha)$, where $N_{\text{pairs}}(\alpha)$ is the
number of base pairs. This model is known in the literature as the maximum
matching model or the Nussinov algorithm \cite{nussinov1978}. It is the simplest
combinatorial model of RNA secondary structure, in which energy is determined
solely by the number of formed pairs, and all complementary pairs are
considered energetically equivalent. The lowest energy corresponds to the
maximum number of pairings. Two topological constraints are imposed on the
structure \cite{nussinov1978,zuker1981}. First, at least three other bases must
lie between paired bases: $j-i\ge4$ for a pair $(i,j)$. This condition is
necessary for the formation of a hairpin loop \cite{uhlenbeck1973,groebe1988}.
Second, pseudoknots are forbidden: for any two pairs $(i,j)$ and $(k,l)$ with
$i<j$ and $k<l$, a structure where $i<k<j<l$ is forbidden \cite{pleij1985,westhof1997}.
Thus, only non-crossing pairs ($i<j<k<l$) and nested pairs ($i<k<l<j$) are
allowed. These rules make the structure nested, like proper bracket sequences,
enabling the use of recursive algorithms \cite{hofacker1994,mccaskill1990,zuker2003}.
Unlike the approach in \cite{higgs1996}, we apply a direct combinatorial method
to generate structures with the maximum number of pairs and use a direct
criterion for classifying triangles with specified accuracy parameters
\cite{zubarev2026}. The quantitative measure of ultrametricity is the fraction
of nontrivially ultrametric triangles (the degree of ultrametricity).
Based on the obtained data, we test the hypothesis that the degree of
nontrivial ultrametricity is a stable characteristic of the sequence.
Additionally, using directed search methods, we investigate how high a degree
of ultrametricity model RNAs can in principle achieve, and compare the
properties of natural and synthetic sequences.

The paper is organized as follows. Section 2 describes the model of secondary
structure states and introduces an integer metric. Section 3 contains the
definition of the direct ultrametricity criterion, including the concepts of
trivial and nontrivial ultrametricity. Section 4 describes the algorithm we
use for uniform generation of structures with the maximum number of pairs with
control over the number of structures. Section 5 presents the results of
numerical experiments on real reference RNAs. Section 6 formulates and solves
the optimization problem for the degree of ultrametricity under a fixed
nucleotide composition, and presents synthetic sequences with extreme
properties. Section 7 contains a discussion of the obtained results,
limitations of the model, and prospects for further research.

\section{Model of RNA Secondary Structure States and Metric}

\label{sec_model}

In the model we consider, the state space $\Omega$ of RNA secondary
structures is defined as the set of all admissible secondary structures for a
given sequence of length $N$. Each secondary structure $\alpha\in\Omega$ is
encoded by an integer vector $b^{\alpha}=(b_{1}^{\alpha},\ldots,b_{N}^{\alpha})$,
where $b_{i}^{\alpha}=0$ if base $i$ is unpaired, and $b_{i}^{\alpha}=j$,
if base $i$ is paired with base $j$. From the single-pairing condition, it
follows that if $b_{i}^{\alpha}=j$, then $b_{j}^{\alpha}=i$.
Among all structures $\Omega$, denote by $\Omega_{\max}$ the subset
consisting of structures with the maximum number of pairs \cite{higgs1996}.
Also denote $P_{\max}=\max_{\alpha\in\Omega}N_{\text{pairs}}(\alpha)$.
Then
\[
\Omega_{\max}=\{\alpha\in\Omega:N_{\text{pairs}}(\alpha)=P_{\max}\}.
\]
In this work, we investigate the ultrametric properties of precisely this
subset $\Omega_{\max}$. All structures in $\Omega_{\max}$ have the same energy
and, consequently, equal statistical weights at $T=0$. At $T=0$, the Boltzmann
distribution is degenerate: all states with the maximum number of pairs have
the same probability, and the rest have zero probability. Therefore, the task
of constructing a representative subset of ground structures reduces to uniform
sampling from $\Omega_{\max}$.

The recurrence relations for calculating the maximum number of pairs, given
below, exactly correspond to the classical Nussinov algorithm \cite{nussinov1978},
adapted not only to count the maximum number of pairs but also the number of
optimal structures. Specifically, we calculate both the maximum number of
nucleotide pairs that can be formed on a given segment and the number of
different secondary structures on that segment, which subsequently allows us
to generate a uniform sample from the set $\Omega_{\max}$.

It is important to emphasize that the maximum pairing model (Nussinov model)
we use is a limiting case where all complementary pairs are considered
energetically equivalent. More realistic thermodynamic models (e.g., the Turner
model) account for differences in the energies of AU and GC pairs, stacking
energies, and loop penalties, which typically lifts the degeneracy.
Nevertheless, the hierarchical structure of the energy landscape is expected
to persist in more complex models, and the quantitative values of the degree
of ultrametricity obtained in this work should be interpreted as a
characteristic of the combinatorial geometry of the set of maximum pairings.
The ensemble $\Omega_{\max}$ at $T=0$ in this degenerate model represents an
idealized object, allowing the study of the geometry of energy minima without
the influence of thermal fluctuations.

In Higgs's work \cite{higgs1996}, the overlap between two structures
$\alpha$ and $\beta$ was defined as the fraction of positions where the pairing
status is the same:
\begin{equation}
q^{\alpha\beta}=\frac{1}{N}\sum_{i=1}^{N}\delta(b_{i}^{\alpha},b_{i}^{\beta}),\label{q_higgs}
\end{equation}
where $\delta(x,y)=1$ if $x=y$ and $0$ otherwise. The distance
between structures was defined as
\begin{equation}
d_{\text{H}}^{\alpha\beta}=1-q^{\alpha\beta}.\label{d_higgs}
\end{equation}
Metric (\ref{d_higgs}) has a simple meaning: it compares the vectors
$b^{\alpha}$ and $b^{\beta}$ elementwise. If $b_{i}^{\alpha}=b_{i}^{\beta}=0$,
then both bases are unpaired and the corresponding contribution to sum (\ref{q_higgs})
is 1. If $b_{i}^{\alpha}=j$ and $b_{i}^{\beta}=j$ (i.e., both bases are paired
with the same partner $j$), the contribution to sum (\ref{q_higgs}) is also 1.
An alternative approach is to compare not the vectors $b_{i}$ but the sets of
pairs. Each structure $\alpha$ is given by a set of unordered pairs:
\[
P_{\alpha}=\bigl\{\{i,j\}:\text{bases }i\text{ and }j\text{ are paired in structure }\alpha\bigr\}.
\]
Following \cite{levandowsky1971}, we can define the Jaccard index between two
unordered pair sets $P_{\alpha}$ and $P_{\beta}$:
\begin{equation}
J^{\alpha\beta}=\frac{|P_{\alpha}\cap P_{\beta}|}{|P_{\alpha}\cup P_{\beta}|},\label{J}
\end{equation}
where $|\cdot|$ denotes cardinality. If both sets are empty,
we set $J=1$. Using (\ref{J}), we can define the Jaccard metric \cite{levandowsky1971}:
\begin{equation}
d_{\text{J}}^{\alpha\beta}=1-J^{\alpha\beta}.\label{d_J}
\end{equation}
For structures with the maximum number of pairs, all structures have the same
number of pairs $|P_{\alpha}|=P_{\max}$, so formula (\ref{d_J}) simplifies:
\[
d_{\text{J}}^{\alpha\beta}=1-\frac{|P_{\alpha}\cap P_{\beta}|}{2P_{\max}-|P_{\alpha}\cap P_{\beta}|}=\frac{2P_{\max}-2|P_{\alpha}\cap P_{\beta}|}{2P_{\max}-|P_{\alpha}\cap P_{\beta}|}.
\]

Both metrics (\ref{d_higgs}) and (\ref{d_J}) are valid.
Nevertheless, it should be noted that metric (\ref{d_higgs}) accounts for
coincidences of unpaired bases. However, unpaired bases do not carry
independent information, since they are completely determined by the set of
pairs. Indeed, knowing all pairs, we know which bases are paired and which are
not. Therefore, although accounting for unpaired bases in metric (\ref{d_higgs})
does not add new information, it can affect the numerical values of distances.

In this work, we introduce an integer metric based directly on the size of the
intersection of the sets of pairs:
\begin{equation}
d^{\alpha\beta}=P_{\max}-|P_{\alpha}\cap P_{\beta}|.\label{d_intersect}
\end{equation}
This metric equals the number of pairs in the maximum structure minus the number
of common pairs between two structures. Since $|P_{\alpha}|=P_{\max}$
for all $\alpha\in\Omega_{\max}$, the quantity $d^{\alpha\beta}$ takes integer
values from $0$ (identical structures) to $P_{\max}$ (structures sharing no
common pair). Metric (\ref{d_intersect}) is closely related to the Jaccard
metric, which further justifies its use. Indeed, using the relations
$|P_{\alpha}\cap P_{\beta}|=P_{\max}-d^{\alpha\beta}$ and
$|P_{\alpha}\cup P_{\beta}|=2P_{\max}-|P_{\alpha}\cap P_{\beta}|=P_{\max}+d^{\alpha\beta}$,
we obtain:
\begin{equation}
d_{\text{J}}^{\alpha\beta}=1-\frac{P_{\max}-d^{\alpha\beta}}{P_{\max}+d^{\alpha\beta}}=\frac{2d^{\alpha\beta}}{P_{\max}+d^{\alpha\beta}}.\label{dJ_from_d}
\end{equation}
Conversely, the integer metric is expressed via the Jaccard distance as
\begin{equation}
d^{\alpha\beta}=P_{\max}\cdot\frac{d_{\text{J}}^{\alpha\beta}}{1-d_{\text{J}}^{\alpha\beta}/2}.\label{d_from_dJ}
\end{equation}
Thus, metric (\ref{d_intersect}) is a simple integer alternative to the Jaccard
metric and allows comparing structures while ignoring unpaired bases. This
metric is used for all subsequent analyses.

\section{Ultrametricity Criterion}

\label{sec_ultr}

The ultrametric property of a metric space $M$, consisting of points
$\left\{ \alpha\right\} $, means that $\forall\:\alpha,\beta,\gamma\in M$,
the strong triangle inequality holds \cite{rammal1986}:
\[
d^{\alpha\gamma}\le\max(d^{\alpha\beta},d^{\beta\gamma}).
\]
An equivalent formulation: in any triangle formed by three points, the two
largest sides are equal. If in a fixed triangle
$\left(\alpha,\beta,\gamma\right)$ we order the side lengths:
$d_{\min}=\min\left(d^{\alpha\beta},d^{\beta\gamma},d^{\alpha\gamma}\right)$,
$d_{\max}=\max\left(d^{\alpha\beta},d^{\beta\gamma},d^{\alpha\gamma}\right)$,
$d_{\text{mid}}$ is the length of the remaining side, then the ultrametricity
condition is written as $d_{\text{mid}}=d_{\max}$.

As noted above, Higgs \cite{higgs1996} proposed an ultrametricity criterion
consisting of the following. All triangles are iterated over and
$s=\langle d_{\max}-d_{\text{mid}}\rangle$ is calculated --- the average
difference between the largest and middle sides of the triangles. Then
uncorrelated triangles are generated: from all available pairwise distances, a
set of triples of numbers is randomly chosen, and for each triple the triangle
inequality is checked. Then all triples satisfying the triangle inequality are
selected, and on this set of triples the average difference between the largest
and middle numbers is calculated, denoted $s_{\text{unc}}$.
If $s/s_{\text{unc}}<1$, the conclusion is that ultrametricity is present.
According to this criterion, the presence of ultrametricity in a real metric
space means that in real triangles the difference between the largest and
middle sides is, on average, smaller than in random triangles constructed from
the same distances. Nevertheless, it should be noted that when testing
ultrametricity, one should distinguish between "trivial ultrametricity" and
"nontrivial ultrametricity". Trivial ultrametricity occurs when the distances
between any three points in a metric space are equal. Such ultrametricity can
arise in systems without real hierarchy, e.g., when randomly choosing points in
a high-dimensional space \cite{zubarev2014,zubarev2017}.
The criterion $s/s_{\text{unc}}<1$ does not distinguish between trivial and
nontrivial ultrametricity. Furthermore, a situation is possible where most
triangles are strictly ultrametric, i.e., they satisfy the exact equality
$d_{\text{mid}}=d_{\max}$, but the remaining non-ultrametric triangles may have
an extremely large difference between $d_{\text{mid}}$ and $d_{\max}$, which
could lead to the ratio $s/s_{\text{unc}}>1$, despite the prevalence of
ultrametric triangles. This is precisely the situation we will observe in some
numerical experiments described below. Finally, the Higgs criterion does not
provide intuitive information about what fraction of triangles actually satisfy
the condition $d_{\text{mid}}=d_{\max}$ and with what accuracy this condition
should be met.

For the reasons listed above, in this work we use the ultrametricity criterion
\cite{zubarev2026} based on the classification of each triangle with a given
accuracy, which we call the "direct ultrametricity criterion". The main
indicator of this criterion is the degree of nontrivial ultrametricity
$u_{\varepsilon,\delta}$, defined below. For formal exposition, we introduce
the following definitions. A triangle will be called trivially ultrametric if
$d_{\min}=d_{\text{mid}}=d_{\max}$ (all three sides equal). A triangle will be
called nontrivially ultrametric if $d_{\text{mid}}=d_{\max}$ and
$d_{\min}<d_{\text{mid}}$ (the two largest sides are equal, and the smallest is
strictly smaller). A triangle that is neither trivially nor nontrivially
ultrametric will be called non-ultrametric. The quantitative measure of
ultrametricity is the "degree of ultrametricity", which equals the fraction of
nontrivially ultrametric triangles, expressed as a percentage. Let
$\mathcal{T}$ be the set of all unordered triples of distinct elements from
$M$. The cardinality of this set is
\[
|\mathcal{T}|=\binom{|M|}{3}.
\]
Let $\mathcal{T}_{\text{nt}}\subset\mathcal{T}$ be the subset of
nontrivially ultrametric triangles from $M$. The degree of nontrivial
ultrametricity of the space $M$ is defined as
\begin{equation}
u=\frac{|\mathcal{T}_{\text{nt}}|}{|\mathcal{T}|}\times100\%,\label{u}
\end{equation}
i.e., the fraction of nontrivially ultrametric triangles among all triangles,
expressed as a percentage. This criterion allows a quantitative assessment of
the degree of ultrametricity in different metric spaces.
Furthermore, let two numbers $\varepsilon$ and $\delta$ be given ($\delta>\varepsilon>0$).
A triangle is considered approximately trivially ultrametric with accuracy
$\varepsilon$ if $(d_{\max}-d_{\min})/d_{\min}\leq\varepsilon$.
A triangle is considered approximately nontrivially ultrametric with accuracy
$\left(\varepsilon,\:\delta\right)$ if $(d_{\max}-d_{\text{mid}})/d_{\text{mid}}\leq\varepsilon$
and $(d_{\text{mid}}-d_{\min})/d_{\text{mid}}>\delta$. In other cases,
a triangle with accuracy $\left(\varepsilon,\:\delta\right)$
is considered non-ultrametric. In this case, we can define the degree of
nontrivial ultrametricity with a given accuracy $\left(\varepsilon,\:\delta\right)$
as follows:
\begin{equation}
u_{\varepsilon,\delta}=\frac{|\mathcal{T}_{\text{nt},\varepsilon,\delta}|}{|\mathcal{T}|}\times100\%,\label{u_epsilon_delta}
\end{equation}
where $\mathcal{T}_{\text{nt},\varepsilon,\delta}\subset\mathcal{T}$
is the subset of triangles from $M$ that are nontrivially ultrametric with
accuracy $\left(\varepsilon,\:\delta\right)$. Obviously, $u_{\varepsilon,\delta}\geq u$
and $u_{0,0}=u$.

\section{Algorithm for Generating Structures with the Maximum Number of Pairs in the RNA Secondary Structure Model}

\label{sec_alg}

In this section, we describe the algorithm used for the numerical generation of
structures with the maximum number of pairs with a strictly uniform
distribution. The algorithm is based on the simultaneous recursive calculation
of two matrices: the maximum number of pairs $M_{ij}$ and the number of optimal
structures $N_{ij}$ on each interval $[i,j]$, followed by probabilistic choice
of pairing variants proportional to the number of achievable optimal
structures. The recurrence relations for $M_{ij}$ exactly correspond to the
classical Nussinov algorithm \cite{nussinov1978}, and the addition of $N_{ij}$
subsequently allows uniform sampling.

Let $N$ be the length of the sequence (the number of nucleotides in the RNA).
Nucleotides are numbered $1,2,3,\ldots,N$. Denote by $[i,j]$ the segment of the
chain from nucleotide $i$ to nucleotide $j$ inclusive. Let $M_{ij}$ be the
maximum number of nucleotide pairs that can be formed on segment $[i,j]$ while
respecting all rules (a pair must be complementary, the distance between bases
in a pair must be at least 4, pseudoknots are forbidden). Let $N_{ij}$ be the
number of different secondary structures on segment $[i,j]$ that achieve this
maximum number of pairs (optimal structures). Both matrices are calculated
simultaneously using the recurrence formulas given below.

For initialization, we set for all $i$: $M_{i,i-1}=0$, $N_{i,i-1}=1$
(an empty segment contains exactly one structure --- the absence of bases);
$M_{ii}=0$, $N_{ii}=1$ (a segment of one unpaired base also gives exactly one
structure). Next, for each interval $[i,j]$ of increasing length $L=j-i+1$, all
admissible structures are considered. Base $j$ can be either unpaired or paired
with some base $h$, where $i\le h\le j-4$ and bases $h$ and $j$ are
complementary. The maximum number of pairs is determined by the relation:
\begin{equation}
M_{ij}=\max\Bigl(M_{i,j-1},\;\max_{\substack{i\le h\le j-4\\
\text{allowed}[h][j]
}
}\bigl[1+M_{i,h-1}+M_{h+1,j-1}\bigr]\Bigr).\label{M_recursive}
\end{equation}
The number of optimal structures $N_{ij}$ is calculated as the sum of the
number of structures for all variants that realize the value $M_{ij}$:
\[
N_{ij}=\begin{cases}
N_{i,j-1}, & \text{if }M_{i,j-1}=M_{ij},\\[2pt]
0, & \text{otherwise}
\end{cases}\;+\!\sum_{\substack{i\le h\le j-4\\
\text{allowed}[h][j]\\[2pt]
1+M_{i,h-1}+M_{h+1,j-1}=M_{ij}
}
}N_{i,h-1}\times N_{h+1,j-1}.
\]
Here $N_{i,h-1}\times N_{h+1,j-1}$ is the number of ways to independently
choose optimal structures on the left and right subsegments resulting from the
formation of pair $(h,j)$ on interval $[i,j]$. Note that the values $N_{ij}$ can
be very large, and Python's long integer arithmetic is used in the software
implementation to handle them. For sequence lengths of 100--200 nucleotides,
the $N_{ij}$ values reach $10^{20}$--$10^{50}$, which does not lead to overflow.

After calculating the matrices $M_{ij}$ and $N_{ij}$, a specific structure is
generated. First, we consider a chain segment from some base $i$ to some base
$j$, where $1\le i\le j\le N$. Our goal is to fill the array elements
$b_{i},b_{i+1},\ldots,b_{j}$ such that the total number of pairs on this
segment equals the precomputed value $M_{ij}$. Initially, all array elements
$b_{i},b_{i+1},\ldots,b_{j}$ are set to zero. We denote the procedure for
filling $b_{i},b_{i+1},\ldots,b_{j}$ by the operator $\hat{F}\left(i,j\right)$,
which is a random mapping of the set of all possible arrays
$\left\{ \left(b_{i},b_{i+1},\ldots,b_{j}\right)\right\} $ to itself.
The recursive action of the operator $\hat{F}(i,j)$ fills the array
$b_{i},b_{i+1},\ldots,b_{j}$, choosing at each step one of the admissible
pairing options with a probability proportional to the number of optimal
structures achievable through that option. More specifically, to define the
action of $\hat{F}(i,j)$ on the array $b_{i},b_{i+1},\ldots,b_{j}$, we form the
set of pairing options on the segment $[i,j]$ as the union of two options.
In the first option, $j$ is unpaired and $M_{i,j-1}=M_{ij}$.
In the second option, there exists $h\in[i,j-4]$ for which $j$ is paired with
$h$ (bases $h$ and $j$ are complementary) and in this case
$1+M_{i,h-1}+M_{h+1,j-1}=M_{ij}$.
For each option, its statistical weight is defined:
\[
\omega_{0}=N_{i,j-1}
\]
in the first option and
\[
\omega_{h}=N_{i,h-1}\cdot N_{h+1,j-1},\;h=i,\ldots,j-4
\]
in the second option. Then an option is chosen randomly with probability
proportional to its weight:
\[
P_{m}=\frac{\omega_{k}}{\omega_{0}+\sum_{k=i}^{j-4}\omega_{k}},
\]
where $m=0$ corresponds to the first option and $m=i,\ldots,j-4$ corresponds to
the second option. After choosing a specific option, the action of the operator
$\hat{F}$ is recursively applied to the corresponding subsegments. If the first
option is chosen, we set $b_{j}=0$ and apply the operator $\hat{F}(i,j-1)$ to the
array $\left(b_{i},b_{i+1},\ldots,b_{j-1}\right)$. If the second option for a
specific $h$ is chosen, we set $b_{h}=j$, $b_{j}=h$ and then apply the operators
$\hat{F}(i,h-1)$ and $\hat{F}(h+1,j-1)$ to the arrays
$\left(b_{i},\ldots,b_{h-1}\right)$ and $\left(b_{h+1},\ldots,b_{j-1}\right)$,
respectively.

The correctness of this algorithm is guaranteed by the fact that for each
interval the sum of weights of all options exactly equals $N_{ij}$, and the
recursive application of proportional selection ensures that each of the
$N_{1N}$ optimal structures of the entire chain is generated with equal
probability $1/N_{1N}$. Thus, the sample is strictly uniform over the set
$\Omega_{\max}$.

\section{Results of Numerical Experiments on Real Reference RNAs}

\label{sec_res}

For each sequence, $M=400$ structures were generated using the algorithm
described in Section \ref{sec_alg}. Duplicates (multiple structures with an
identical set of pairs) that appeared during generation were removed, so that
only unique structures were included in the final sample. Thus, the effective
sample size $M_{\text{eff}}$ for each sequence could be less than 400 and was
determined by the number of different generated structures. For statistical
robustness, 10 independent realizations of a random sample of a given size from
$\Omega_{\max}$ were used for each sequence (with different initial seeds of
the random number generator). For each realization, the full matrix of pairwise
distances between all unique structures was calculated using the integer metric
(\ref{d_intersect}), followed by classification of all $\binom{M_{\text{eff}}}{3}$
triangles according to the direct criterion with parameters $(\varepsilon=0.05,\ \delta=0.05)$
and $(\varepsilon=0,\ \delta=0)$. For each sequence, averaging over the 10
realizations gave the mean value of the degree of ultrametricity and the
standard deviation for both sets of parameters.

To test the Higgs ultrametricity criterion, the value $s/s_{\text{unc}}$ was
additionally calculated for the same data. The value $s$ was determined as the
average value of $d_{\max}-d_{\text{mid}}$ over all real triangles. To compute
$s_{\text{unc}}$, triples of numbers were randomly selected from all pairwise
distances, the triangle inequality was checked, and the average difference
between the largest and middle numbers was computed for the selected triples.
For statistical stability, 10,000 random attempts were used.

The results of numerical experiments for 18 reference small nuclear RNAs from
the NCBI database \cite{refseq} are presented in Table \ref{tab:nat_rna}.
For each RNA, the full name, length in nucleotides, degree of exact nontrivial
ultrametricity $u_{0,0}$ (main indicator), fraction of trivial (equilateral)
triangles, degree of approximate ultrametricity $u_{0.05,0.05}$, and the value
of the indicator $s/s_{\text{unc}}$ (provided for comparability with \cite{higgs1996}) are given.

\begin{table}[H]
\centering {\small{}}{\small{}%
\begin{tabular}{p{4.0cm}p{1cm}p{2.2cm}p{1.3cm}p{2.2cm}p{1.3cm}}
\toprule
{\small RNA  } & {\small Length (nt)  } & {\small$u_{0,0}$ (\%)  } & {\small Triv. (\%)  } & {\small$u_{0.05,0.05}$ (\%)  } & {\small$s/s_{\text{unc}}$ }\tabularnewline
\midrule
{\small Leishmania major U5  } & {\small 72  } & {\small$25.25\pm0.25$  } & {\small 2.25  } & {\small$25.62\pm0.34$  } & {\small 0.7377 }\tabularnewline
{\small C. elegans smy-7  } & {\small 82  } & {\small$13.77\pm0.17$  } & {\small 0.57  } & {\small$15.30\pm0.43$  } & {\small 0.7513 }\tabularnewline
{\small A. thaliana U5 (Ath-134a)  } & {\small 94  } & {\small$18.02\pm0.14$  } & {\small 0.49  } & {\small$18.02\pm0.14$  } & {\small 0.6751 }\tabularnewline
{\small C. elegans T10H4.17  } & {\small 95  } & {\small$12.92\pm0.11$  } & {\small 0.48  } & {\small$16.02\pm0.67$  } & {\small 0.6917 }\tabularnewline
{\small C. elegans sls-1.10  } & {\small 98  } & {\small$35.04\pm0.36$  } & {\small 1.85  } & {\small$35.04\pm0.36$  } & {\small 0.9518 }\tabularnewline
{\small C. elegans B0213.24  } & {\small 102  } & {\small$13.92\pm0.19$  } & {\small 0.56  } & {\small$14.89\pm0.21$  } & {\small 0.7247 }\tabularnewline
{\small C. elegans F32D1.12  } & {\small 102  } & {\small$45.94\pm1.64$  } & {\small 0.49  } & {\small$60.36\pm2.00$  } & {\small 0.8189 }\tabularnewline
{\small C. elegans ZK1251.14  } & {\small 104  } & {\small$11.62\pm0.13$  } & {\small 0.26  } & {\small$14.18\pm0.31$  } & {\small 0.8582 }\tabularnewline
{\small C. elegans F35B3.10  } & {\small 104  } & {\small$13.24\pm0.24$  } & {\small 0.24  } & {\small$31.00\pm0.78$  } & {\small 0.5836 }\tabularnewline
{\small C. elegans F56A12.5  } & {\small 108  } & {\small$8.39\pm0.11$  } & {\small 0.16  } & {\small$17.51\pm0.23$  } & {\small 0.8891 }\tabularnewline
{\small Leishmania infantum U4  } & {\small 116  } & {\small$22.34\pm0.15$  } & {\small 1.89  } & {\small$22.34\pm0.15$  } & {\small 0.8991 }\tabularnewline
{\small C. elegans Y69A2AR.36  } & {\small 121  } & {\small$14.38\pm0.33$  } & {\small 0.32  } & {\small$23.87\pm0.39$  } & {\small 0.7235 }\tabularnewline
{\small A. thaliana U2 (Ath-227c)  } & {\small 131  } & {\small$22.94\pm0.54$  } & {\small 0.51  } & {\small$49.92\pm1.15$  } & {\small 0.7324 }\tabularnewline
{\small Leishmania braziliensis U2  } & {\small 134  } & {\small$11.83\pm0.12$  } & {\small 0.39  } & {\small$13.70\pm0.16$  } & {\small 0.7993 }\tabularnewline
{\small S. osmophilus U6  } & {\small 155  } & {\small$9.55\pm0.19$  } & {\small 0.22  } & {\small$25.41\pm0.62$  } & {\small 0.6001 }\tabularnewline
{\small C. elegans T27E4.22  } & {\small 156  } & {\small$17.85\pm0.42$  } & {\small 0.15  } & {\small$36.35\pm1.09$  } & {\small 0.4557 }\tabularnewline
{\small C. elegans C05G6.4  } & {\small 186  } & {\small$7.24\pm0.06$  } & {\small 0.10  } & {\small$25.24\pm0.39$  } & {\small 0.6618 }\tabularnewline
{\small S. osmophilus U2  } & {\small 191  } & {\small$14.64\pm0.24$  } & {\small 0.72  } & {\small$20.23\pm0.59$  } & {\small 0.5840 }\tabularnewline
\bottomrule
\end{tabular}}{\small\caption{Results of ultrametricity analysis for 18 reference small nuclear RNAs. The main indicator is the degree of exact nontrivial ultrametricity $u_{0,0}$. The fraction of trivial triangles (Triv.) is small in all cases ($<2.5\%$). The indicators $u_{0.05,0.05}$ and $s/s_{\text{unc}}$ are provided for comparability with \cite{higgs1996}.}
\label{tab:nat_rna} }
\end{table}

Analysis of the obtained data shows that for most of the studied natural RNAs,
the degree of exact nontrivial ultrametricity does not exceed 25\%.
At the same time, a significant spread of values is observed: from a minimum
of $7.24\%$ (Caenorhabditis elegans C05G6.4 snRNA) to a maximum of $45.94\%$
(Caenorhabditis elegans F32D1.12 snRNA). The fraction of trivial (equilateral)
triangles is small in all cases (less than $2.5\%$), indicating that the
observed ultrametricity is not an artifact of the degenerate case of all equal
distances. Upon introducing the tolerance $\varepsilon=0.05,\delta=0.05$, the
maximum value increases to $60.36\%$ for the same RNA F32D1.12, indicating the
presence of a significant fraction of triangles close to ultrametric but not
satisfying the exact equality of sides. It is noteworthy that for some RNAs
(e.g., Leishmania major U5 snRNA, Caenorhabditis elegans sls-1.10 snRNA,
Leishmania infantum U4 snRNA), the values of $u_{0.05,0.05}$ and $u_{0,0}$ are
consistent within error, indicating that most ultrametric triangles in these
systems are strictly ultrametric. For other RNAs (Caenorhabditis elegans
F35B3.10 snRNA, Arabidopsis thaliana U2 snRNA, Schizosaccharomyces osmophilus
U6 snRNA, Caenorhabditis elegans T27E4.22 snRNA, Caenorhabditis elegans
C05G6.4 snRNA), a sharp increase in the degree of ultrametricity is observed
when transitioning from the strict to the approximate criterion, indicating the
presence of a large number of triangles for which the equality
$d_{\text{mid}}=d_{\max}$ holds only approximately.

It is important to note that the Higgs criterion $s/s_{\text{unc}}$ does not
always correlate with the direct fraction of nontrivially ultrametric triangles.
As will be shown in the next section, for synthetic sequences with a very high
degree of ultrametricity ($80\%$), this criterion yields values
$s/s_{\text{unc}}>1$, which is formally interpreted as the absence of
ultrametricity. Therefore, in this work, the criterion $s/s_{\text{unc}}$ is
provided only for comparability with the results of \cite{higgs1996}, but is not
used for the main conclusions.

Based on visual analysis of the data in Table \ref{tab:nat_rna}, no monotonic
dependence between sequence length and the degree of ultrametricity is observed.
A short RNA of length 72 (Leishmania major U5) demonstrates moderate
ultrametricity ($25.25\%$), whereas an RNA of length 98 (Caenorhabditis elegans
sls-1.10) has a higher value ($35.04\%$), and an RNA of length 186
(Caenorhabditis elegans C05G6.4) has the minimum ($7.24\%$). This indicates
that ultrametric properties are determined not simply by length, but by the
specific arrangement of complementary regions along the sequence.

\section{Search for RNA Sequences with Maximum and Minimum Degrees of Secondary Structure Ultrametricity under Fixed Nucleotide Composition}

\label{sec_max_ultr}

The results presented in the previous section show that the degree of
nontrivial ultrametricity for natural reference RNAs varies widely but does not
reach extremely high values. This raises a natural question: how high (or low)
a degree of ultrametricity can RNA sequences in principle achieve, given the
same nucleotide composition as the natural molecules? To answer this question,
the optimization problem for the degree of nontrivial ultrametricity is
formulated and solved as a discrete optimization problem on the space of
sequences with fixed numbers of each nucleotide type. Discrete optimization
problems of this kind (finding the global extremum of a deterministic function
on a finite set of large cardinality) are discussed in detail in integer
programming and combinatorial optimization theory (see, e.g., \cite{papadimitriou1982,nemhauser1988}).
The approach used in this work, based on iterative probabilistic directed search
with adaptive parameter control, belongs to the class of adaptive random search
algorithms \cite{spall2003}, as well as to directed search methods with learning
\cite{glover1997}.

Let $N\in\mathbb{N}$, $N\ge10$, be the sequence length and $\Sigma=\{A,C,G,U\}$
the nucleotide alphabet. The space of all sequences of length $N$ over the
alphabet $\Sigma$ is the set
\[
X_{N}=\Sigma^{N}=\{s=(s_{1},s_{2},\ldots,s_{N})\mid s_{i}\in\Sigma,\;i=1,2,\ldots,N\}.
\]
Let a sequence $s_{0}\in X_{N}$ be given. Denote by $n_{A},n_{C},n_{G},n_{U}$
the counts of nucleotides $A,C,G,U$ in sequence $s_{0}$. Obviously,
$n_{A}+n_{C}+n_{G}+n_{U}=N$. The set of sequences with fixed nucleotide
composition generated by the sequence $s_{0}$ is the set $X_{N}(s_{0})$,
consisting of all sequences obtained by permuting the symbols of sequence
$s_{0}$, i.e.,
\[
X_{N}(s_{0})=\{s\in X_{N}\mid\forall\,x\in\Sigma:\;|\{i\mid s_{i}=x\}|=|\{i\mid(s_{0})_{i}=x\}|\}.
\]
The cardinality of this set is
\[
|X_{N}(s_{0})|=\frac{N!}{n_{A}!\,n_{C}!\,n_{G}!\,n_{U}!}.
\]
The set $X_{N}(s_{0})$ is the search space in the considered problem.

For a sequence $s\in X_{N}(s_{0})$ and the corresponding metric space
$(\Omega_{\max}(s),d_{s})$, where the metric $d_{s}$ is defined by formula
(\ref{d_intersect}) with $P_{\max}$ replaced by $P_{\max}(s)$, we define the
objective function. Let $\mathcal{T}(s)$ be the set of all unordered triples of
distinct elements of $\Omega_{\max}(s)$:
\[
\mathcal{T}(s)=\{\{\alpha,\beta,\gamma\}\subset\Omega_{\max}(s)\mid\alpha\neq\beta,\;\beta\neq\gamma,\;\alpha\neq\gamma\}.
\]
Let $\mathcal{T}_{\text{nt}}(s)\subset\mathcal{T}(s)$ be the subset of
nontrivially ultrametric triangles (in the sense of the definition with
parameters $\varepsilon=0$, $\delta=0$, given in Section \ref{sec_ultr}).
Then the degree of nontrivial ultrametricity of sequence $s$ is
\[
u(s)=\frac{|\mathcal{T}_{\text{nt}}(s)|}{|\mathcal{T}(s)|}\times100\%.
\]
The function $u:X_{N}(s_{0})\to[0,100]$ is deterministic: each sequence $s$ is
assigned exactly one number, completely determined by the set $\Omega_{\max}(s)$.
If $|\Omega_{\max}(s)|<3$, we set $u(s)=0$ by definition.

Next, we consider two optimization problems on the set $X_{N}(s_{0})$.
The first problem is maximizing the degree of nontrivial ultrametricity under a
fixed nucleotide composition. It is defined as finding a sequence
$s^{*}\in X_{N}(s_{0})$ such that $u(s^{*})=\max_{s\in X_{N}(s_{0})}u(s)$.
The second problem is minimizing the degree of nontrivial ultrametricity under
a fixed nucleotide composition. This is the problem of finding a sequence
$s^{*}\in X_{N}(s_{0})$ such that $u(s^{*})=\min_{s\in X_{N}(s_{0})}u(s)$.
These problems are classical discrete optimization problems with a deterministic
objective function on a permutation space.

To solve these problems, an iterative probabilistic directed search algorithm
is applied. The algorithm works with a finite subset $\mathcal{P}\subset X_{N}(s_{0})$
of fixed size $M\in\mathbb{N}$, $M\ge2$, called the current set of elements.
The initial set $\mathcal{P}_{0}$ consists of $M$ elements of the space
$X_{N}(s_{0})$, each generated as a random equiprobable permutation of the
symbols of sequence $s_{0}$. At each step $t=0,1,2,\ldots$, the current set
$\mathcal{P}_{t}$ is transformed into a set $\mathcal{P}_{t+1}$ of the same size
$M$. For this, the operators defined below are used. All operators are
constructed so that their image lies in $X_{N}(s_{0})$, i.e., the nucleotide
composition is preserved.

The operator preserving the best elements $\hat{P}_{E}$ is defined as follows.
Fix an integer $E$ such that $1\le E<M$.
The operator $\hat{P}_{E}$ maps each current set $\mathcal{P}_{t}\subset X_{N}(s_{0})$,
$|\mathcal{P}_{t}|=M$, to its subset $\hat{P}_{E}(\mathcal{P}_{t})\subset\mathcal{P}_{t}$,
containing exactly $E$ elements. Namely, order the elements of $\mathcal{P}_{t}$
in non-increasing order of $u$ values (for the maximization problem) or
non-decreasing order (for the minimization problem):
\[
u(s_{(1)})\ge u(s_{(2)})\ge\cdots\ge u(s_{(M)}).
\]
Then
\[
\hat{P}_{E}(\mathcal{P}_{t})=\left\{ s_{(1)},s_{(2)},\ldots,s_{(E)}\right\} .
\]

Next, define the selection operator $\hat{S}_{k}$. Fix an integer $k$ such that
$2\le k\le M$. The operator $\hat{S}_{k}$ maps the current set
$\mathcal{P}_{t}\subset X_{N}(s_{0})$ to one element $s\in X_{N}(s_{0})$
according to the following rule. From $\mathcal{P}_{t}$, a subset
$\mathcal{T}_{k}(\mathcal{P}_{t})$ of size $k$ is drawn randomly and uniformly
(without replacement). The result is the element of this subset with the maximum
value of $u$ (for the maximization problem) or the minimum (for the minimization
problem):
\[
\hat{S}_{k}(\mathcal{P}_{t})=\arg\max_{s\in\mathcal{T}_{k}(\mathcal{P}_{t})}u(s)\;\text{(maximization)},
\]
\[
\hat{S}_{k}(\mathcal{P}_{t})=\arg\min_{s\in\mathcal{T}_{k}(\mathcal{P}_{t})}u(s)\;\text{(minimization)}.
\]
If the extremum is attained by several elements, the choice among them is uniform.

We also define the permutation local change operator $\hat{V}_{q}$. Fix a number
$q\in[0,1]$, called the local change intensity. The operator $\hat{V}_{q}$ maps
an element $s\in X_{N}(s_{0})$ to a new element $s'\in X_{N}(s_{0})$ according
to the following rule. Compute the number of swaps $w=\max(1,\lfloor q\cdot N\rfloor)$.
Then the following procedure is repeated $w$ times: from the set of indices
$\{1,\ldots,N\}$, a pair of distinct indices $(i,j)$ is drawn uniformly and
without replacement, and the values $s_{i}$ and $s_{j}$ are swapped. The element
obtained after $w$ swaps is denoted $\hat{V}_{q}(s)$.

Finally, we define the composition operator preserving composition $\hat{C}$.
Fix a number $p_{c}\in[0,1]$, called the composition probability.
The operator $\hat{C}$ maps a pair of elements $(a,b)\in X_{N}(s_{0})\times X_{N}(s_{0})$
to a new pair $(a',b')\in X_{N}(s_{0})\times X_{N}(s_{0})$ according to the
ordered crossover rule. With probability $p_{c}$, a cut point $\theta$ is chosen,
uniformly distributed on $\{1,\ldots,N-1\}$. The elements $a'$ and $b'$ are
constructed as follows. For $i\in[1,\theta]$, set $a'_{i}=a_{i}$, $b'_{i}=b_{i}$.
For $i\in[\theta+1,N]$, the positions of $a'$ are filled with the remaining
symbols from $b$ in the order of their appearance in $b$ that were not used in
the initial segment of $a'$. Similarly, the positions of $b'$ are filled with
the remaining symbols from $a$ that were not used in the initial segment of $b'$.
With probability $1-p_{c}$, set $(a',b')=(a,b)$ (identity mapping).

Using the introduced operators, the step-by-step formation of the next set
$\mathcal{P}_{t+1}$ from the set $\mathcal{P}_{t}$ is described.

All elements of the subset $\hat{P}_{E}(\mathcal{P}_{t})$ are included in
$\mathcal{P}_{t+1}$. The remaining $M-E$ places in $\mathcal{P}_{t+1}$ are
filled iteratively. Namely, until $|\mathcal{P}_{t+1}|<M$, the following steps
are performed:

1. The selection operator is applied twice: $a=\hat{S}_{k}(\mathcal{P}_{t})$,
$b=\hat{S}_{k}(\mathcal{P}_{t})$.

2. The composition operator preserving composition is applied: $(a',b')=\hat{C}(a,b)$.

3. The permutation local change operator with intensity $q$ is applied to $a'$
and $b'$ separately: $a''=\hat{V}_{q}(a')$, $b''=\hat{V}_{q}(b')$.

4. The elements $a''$ and $b''$ (or only one of them if only one slot remains
to be filled) are included in $\mathcal{P}_{t+1}$.

The local change intensity $q_{t}$ at step $t$ is determined by an adaptive
rule. Let $u_{t}^{*}$ denote the best value of the objective function in the
set $\mathcal{P}_{t}$ (maximum for the maximization problem, minimum for the
minimization problem). Let $\tau_{t}$ be the number of consecutive steps
preceding $t$ during which $u_{t}^{*}$ has not improved by more than a given
small amount $\varepsilon_{u}>0$. If $\tau_{t}=0$ (improvement occurred on the
previous step), then $q_{t}$ is decreased by multiplying by a factor
$\alpha\in(0,1)$:
\[
q_{t+1}=\max(q_{\min},\alpha\cdot q_{t}).
\]
If $\tau_{t}>0$, then $q_{t}$ increases linearly with $\tau_{t}$:
\[
q_{t+1}=\min\bigl(q_{\max},\;q_{\min}+\tau_{t}\cdot\Delta q\bigr),
\]
where $q_{\min}$, $q_{\max}$, $\Delta q$ are fixed parameters satisfying
$0<q_{\min}\le q_{\max}\le1$, $\Delta q>0$. The meaning of this rule is as
follows. When the algorithm successfully finds elements with improving $u$
values, the local change intensity is low, which facilitates detailed
exploration of the neighborhood of the best found elements (local refinement).
When progress stalls, the local change intensity increases, allowing the
algorithm to explore more distant regions of the search space and potentially
escape local extrema.

The iterative process continues until one of the following conditions is met:

- reaching a specified threshold value $u_{\text{target}}\in[0,100]$
(for maximization: $u_{t}^{*}\ge u_{\text{target}}$; for minimization:
$u_{t}^{*}\le u_{\text{target}}$);

- exceeding the maximum number of iterations $T_{\max}\in\mathbb{N}$;

- exceeding the maximum duration without improvement $\tau_{\max}\in\mathbb{N}$.

After completing the iterative process, a sequential refinement procedure is
applied to the best found element $s^{*}$. At each step $i=1,\ldots,I_{\text{ref}}$,
$m_{\text{ref}}$ pairs of distinct indices are drawn uniformly and without
replacement from the set $\{1,\ldots,N\}$, and the corresponding pairs of
symbols are swapped. If the value of the objective function for the obtained
element is better than for $s^{*}$ (higher in the maximization problem, lower
in the minimization problem), then $s^{*}$ is replaced by the new element.

To exclude degenerate solutions (sequences with a small number of structures
with the maximum number of pairs), a threshold value $m_{\min}\in\mathbb{N}$,
$m_{\min}\ge3$, is introduced. An element $s$ is considered admissible only if
$|\Omega_{\max}(s)|\ge m_{\min}$. Inadmissible elements are excluded from
consideration at all stages of the algorithm.

The numerical implementation of the algorithm uses an approximate computation
of the objective function. Exact computation of $u(s)$ would require
enumerating all elements of $\Omega_{\max}(s)$, but $|\Omega_{\max}(s)|$ can
reach values of the order of $10^{20}$ or higher. Therefore, the exact value of
$u(s)$ is replaced by a sample estimate $\check{u}_{n}(s)$ constructed from a
random sample of $n$ elements from $\Omega_{\max}(s)$. The sample is generated
by the algorithm described in Section \ref{sec_alg}, which guarantees a uniform
distribution. Duplicates are removed from the sample, so that it contains
$n_{\mathrm{uniq}}\le n$ unique elements. On the set of unique elements, the
matrix of pairwise distances $d_{s}(\alpha_{p},\alpha_{q})$ is constructed,
then all $\binom{n_{\mathrm{uniq}}}{3}$ triples are enumerated (or a random
sample of triples is used to reduce computations), and for each triple the
condition for nontrivial ultrametricity is checked. The sample estimate is
defined as
\[
\check{u}_{n}(s)=\frac{T_{\mathrm{nt}}}{T_{\mathrm{total}}}\times100\%,
\]
where $T_{\mathrm{nt}}$ is the number of nontrivially ultrametric triples,
and $T_{\mathrm{total}}$ is the total number of triples analyzed.

To increase the reliability of the estimate, the computation is repeated
$K\in\mathbb{N}$ times with independent samples, and the result is averaged:
\[
\bar{u}_{n}(s)=\frac{1}{K_{\text{valid}}}\sum_{j=1}^{K}\check{u}_{n}^{(j)}(s)\cdot\mathbf{1}\{\check{u}_{n}^{(j)}(s)\text{ admissible}\},
\]
where $K_{\text{valid}}$ is the number of admissible estimates (with
$n_{\mathrm{uniq}}\ge m_{\min}$). The variance of $\bar{u}_{n}(s)$ decreases with
increasing $K$ and $n$, making the estimate consistent.

In the numerical experiments conducted, the following parameters of the
optimization algorithm were used: population size $M=40$, number of best
elements $E=8$, size of the drawn set $k=5$, composition probability $p_{c}=0.7$,
initial local change intensity $q_{0}=0.05$, adaptive bounds $q_{\min}=0.01$,
$q_{\max}=0.25$, increase step $\Delta q=0.02$, decrease factor $\alpha=0.9$,
maximum number of iterations $T_{\max}=500$, maximum number of steps without
improvement $\tau_{\max}=50$, improvement threshold $\varepsilon_{u}=0.5\%$,
number of refinement steps $I_{\text{ref}}=50$, number of swaps per refinement
step $m_{\text{ref}}=3$, threshold for minimum number of unique structures
$m_{\min}=20$.

The algorithm was applied to search for sequences with the maximum and minimum
degree of ultrametricity under the fixed nucleotide composition defined by two
reference RNA sequences that showed the minimum and maximum degree of
ultrametricity among the natural samples. As the first reference sequence,
Caenorhabditis elegans C05G6.4 snRNA (length 186 nt, exact ultrametricity
degree $7.24\pm0.06\%$) was chosen. The optimization yielded a sequence with
maximum ultrametricity:

\texttt{RNA C05G6\_4\_Max\_Ultr =}\\
 \texttt{UGGUUCACCG UAUCGAUACC UCGCAUCAGA CUCCCUUCAC UCCUUCCCCC GAAAGGAAAA}\\
 \texttt{GGGAAGAAGG AAGUUCGAGA UGCGCUGUUU UUAAGCUUUU GUCUUAUCUU UUUGCUACGU}\\
 \texttt{CGGAUCGUUG UAUAUUUGUG GGGGGCGCAC CAGUAGGUGA AUACUCUCCA GUAUUAUUCC}\\
 \texttt{AUGAGG}\\
 and a sequence with minimum ultrametricity:

\texttt{RNA C05G6\_4\_Min\_Ultr =}\\
 \texttt{UAUCGCGUGC GCGCCGCGGC CAUUCGCAGC CAUGUUCGAG AGCGGGUGAC GUGACAAACU}\\
 \texttt{CUGACAUAUA GGGCUGCUGU CAAGACUAGC ACUAUCCUUU CAGUGGUAUU UCUAAACUCA}\\
 \texttt{AUGAUGGUAU UGGUCUCCGU CGAUUGGAUC CCGGCCGCGG UUUUAUUUUU UAAAAAAUAU}\\
 \texttt{UUUUAU}.

As the second reference sequence, Caenorhabditis elegans F32D1.12 snRNA
(length 102 nt, exact ultrametricity degree $45.94\pm1.64\%$) was chosen.
For this sequence, permutations with extreme properties were also constructed.
A sequence with maximum ultrametricity was obtained:

\texttt{RNA F32D1\_12\_Max\_Ultr =}\\
 \texttt{AUUUCCCCUC CACCUUGGAA AGAGAGUAAG AGAGGACACA AGUCCCGGUC AUCUCAGCUC}\\
 \texttt{GGGAAAAGAA GUAAAUAUAU CAGGACUUUG UCCUAAUAAA AA}~\\
 and a sequence with minimum ultrametricity:

\texttt{RNA F32D1\_12\_Min\_Ultr =}\\
 \texttt{UGAGCAAAGC UCUGCUUCGA AAGGUUAAUG CUCUCGUCAA CGUGAGACAG AAAAAAAAAA}\\
 \texttt{UCUCUCCAAA UCUCUCUGGG GACGACUCAU CAAAAAGAGA GU}.

To verify the achieved extreme values, these four synthetic sequences were
subjected to a full analysis using the main program. Unlike the optimization
procedure, where the sample estimate $\bar{u}_{n}(s)$ on a subset of structures
was used, the final verification was performed on an *independent* sample of
400 structures (10 realizations) using the same methodology as for the natural
RNAs. This eliminates the effect of overfitting on the sample estimate.
The results are presented in Table \ref{tab:opt_rna}. To assess the robustness
of the results to the random seed, the optimization procedure for sequence
F32D1.12 was repeated three times with different initial seeds.

\begin{table}[H]
\centering {\small{}}{\small{}%
\begin{tabular}{p{4.0cm}p{1cm}p{2.2cm}p{1.3cm}p{2.2cm}p{1.3cm}}
\toprule
{\small RNA  } & {\small Length (nt)  } & {\small$u_{0,0}$ (\%)  } & {\small Triv. (\%)  } & {\small$u_{0.05,0.05}$ (\%)  } & {\small$s/s_{\text{unc}}$ }\tabularnewline
\midrule
{\small C. elegans F32D1.12 (max. optim.)  } & {\small 102  } & {\small$80.94\pm0.71$  } & {\small 0.58  } & {\small$80.94\pm0.71$  } & {\small 1.1932 }\tabularnewline
{\small C. elegans F32D1.12 (min. optim.)  } & {\small 102  } & {\small$2.57\pm0.00$  } & {\small 0.00  } & {\small$3.16\pm0.00$  } & {\small 1.2797 }\tabularnewline
{\small C. elegans C05G6.4 (max. optim.)  } & {\small 186  } & {\small$80.16\pm0.51$  } & {\small 0.70  } & {\small$80.16\pm0.51$  } & {\small 1.2216 }\tabularnewline
{\small C. elegans C05G6.4 (min. optim.)  } & {\small 186  } & {\small$5.42\pm0.14$  } & {\small 0.08  } & {\small$18.63\pm0.29$  } & {\small 0.7378 }\tabularnewline
\bottomrule
\end{tabular}}{\small\caption{Results of ultrametricity analysis for synthetic sequences obtained by optimization under fixed nucleotide composition. The main indicator is the degree of exact nontrivial ultrametricity $u_{0,0}$. For all sequences, the fraction of trivial triangles is small.}
\label{tab:opt_rna} }
\end{table}

These results demonstrate that, under a fixed nucleotide composition, the degree
of nontrivial ultrametricity can vary over an extremely wide range. For an RNA
of length 102 nt (based on F32D1.12), a value of $80.94\%$ was achieved upon
maximization and $2.57\%$ upon minimization. For an RNA of length 186 nt (based
on C05G6.4), the maximum value was $80.16\%$ and the minimum $5.42\%$ under the
strict criterion. Particularly indicative is the case of C05G6.4: the original
natural sequence exhibited one of the lowest degrees of ultrametricity ($7.24\%$),
but by permuting the nucleotides, a sequence with ultrametricity of $80.16\%$,
i.e., more than 11 times higher, was obtained. The opposite situation is
observed for F32D1.12: the original natural sequence had a relatively high
ultrametricity ($45.94\%$), but through optimization it was possible to both
increase it to $80.94\%$ and decrease it to $2.57\%$. It is worth noting that
for the obtained maximized sequences, the values of $u_{0.05,0.05}$ and $u_{0,0}$
coincide within error, indicating that the ultrametricity in them is strict:
the vast majority of triangles classified as nontrivially ultrametric satisfy
the exact equality $d_{\text{mid}}=d_{\max}$. For the minimized sequence based
on F32D1.12, the values are also close ($3.16\%$ and $2.57\%$), indicating that
the small number of ultrametric triangles is strictly ultrametric. For the
minimized sequence based on C05G6.4, a discrepancy is observed ($18.63\%$ vs.
$5.42\%$), indicating the presence of a significant number of triangles close
to ultrametric but not satisfying the exact equality.

The behavior of the Higgs criterion $s/s_{\text{unc}}$ is also interesting.
For the maximized sequences, this indicator exceeds one ($1.19$ and $1.22$),
which formally, according to the Higgs criterion, should be interpreted as the
absence of ultrametricity. However, direct classification shows that more than
$80\%$ of triangles are strictly ultrametric. This contradiction is explained
by the fact that the remaining $\sim20\%$ of non-ultrametric triangles have a
very large difference $d_{\max}-d_{\text{mid}}$, leading to an increase in the
average value $s$ and, consequently, to $s/s_{\text{unc}}>1$. Thus, the Higgs
criterion proves unsuitable for systems with high but not complete
ultrametricity if the non-ultrametric triangles possess extreme properties.

The obtained results convincingly demonstrate that the degree of nontrivial
ultrametricity under a fixed nucleotide composition depends extremely strongly
on the order of nucleotides. This means that the hierarchical organization of
the RNA energy landscape is not unambiguously determined by the composition,
but is a subtle property encoded in the sequence. In this limited sample of
18 sequences, there is a tendency: natural RNAs show intermediate values of
the degree of ultrametricity, not reaching either extremely high or extremely
low values in principle achievable for a given nucleotide composition. This
suggests (although a statistically significant conclusion requires the analysis
of a substantially larger number of sequences) that evolution might have acted
to maintain the degree of ultrametricity within some optimal range, providing a
compromise between folding speed and conformational mobility.

\section{Discussion and Perspectives}

\label{sec_disc}

This study has allowed a quantitative characterization of the ultrametric
properties of the set of structures with the maximum number of base pairs
(energetically degenerate minima) of RNA secondary structures within the
Nussinov model. The use of a direct criterion for classifying triangles with
two sets of accuracy parameters made it possible to distinguish between strict
and approximate ultrametricity, as well as to separate nontrivial ultrametricity
from trivial (equilateral) ultrametricity. The uniform generation algorithm
for structures with the maximum number of pairs, based on the recursive
calculation of the $M_{ij}$ and $N_{ij}$ matrices followed by probabilistic
selection, guarantees the absence of systematic biases in estimating the degree
of ultrametricity. The proposed method for optimizing sequences under a fixed
nucleotide composition made it possible to achieve extreme values of the degree
of ultrametricity and to reveal the fundamental capabilities embedded in a given
nucleotide composition.

Limitations of the presented model and method should be noted. The main
limitation is related to the energy model used. We adopted the simplest scheme
in which each formed pair lowers the energy by one unit, and the structure
energy equals minus the number of pairs. This model, known as the maximum
matching model or the Nussinov algorithm \cite{nussinov1978}, is well studied
and allows an exact solution to the maximum pairing problem, but it does not
account for differences in the stability of different types of pairs (A--U and
G--C), stacking energies, loop penalties, and other thermodynamic factors
present in more realistic models such as the Turner model \cite{freier1986,mathews1999}.
Nevertheless, the choice of this model was motivated by several considerations.
First, it reduces the problem to combinatorial optimization with a well-defined
set of structures having the maximum number of pairs, enabling exact uniform
sampling. Second, it can be expected that the qualitative conclusions about
hierarchical organization will persist in more complex models, since
ultrametricity is a property of the energy landscape, not of the details of
energy parameters. However, the quantitative values of the degree of
ultrametricity obtained in this work should be interpreted as characteristics
of the combinatorial geometry of the set of maximum pairings, and not as a
direct prediction for thermodynamic models. The question of to what extent the
obtained quantitative results (especially the extreme values $u\approx81\%$ for
optimized sequences) will persist when transitioning to the Turner model
remains open and requires separate investigation. It can be expected that in a
more realistic model, where the energies of A--U and G--C pairs differ and
cooperative stacking effects are taken into account, the degree of
ultrametricity may both decrease (due to lifting of degeneracy) and increase
(due to the dominance of G--C-rich pairings). Verification on a limited set of
sequences in the Turner model is planned in the near future. The next limitation
of the model considered is the prohibition of pseudoknots. Pseudoknots do occur
in real RNAs, although they are not ubiquitous. However, including pseudoknots
significantly complicates the combinatorial structure of the problem and makes
it impossible to use simple recursive algorithms like Nussinov's. The problem
of maximizing the number of pairs in the presence of pseudoknots becomes
NP-complete, and exact determination of structures with the maximum number of
pairs for sequences longer than 100 nucleotides becomes practically infeasible.
Therefore, to ensure computational tractability, we limited ourselves to
structures without pseudoknots. This is a standard approximation in RNA
bioinformatics. Another limitation is associated with using only structures
with the maximum number of pairs (corresponding to $T=0$). At finite
temperatures, excited states also contribute to the partition function, and
ultrametric properties may manifest differently. Our approach focuses on the
limit $T\rightarrow0$, allowing the study of the pure geometry of the set of
global pairing maxima without the influence of thermal fluctuations.

Based on the results obtained in this work, a number of conclusions can be
drawn. Natural small nuclear RNAs demonstrate a significant spread in the
degree of ultrametricity (from 7\% to 46\% under the strict criterion).
This indicates that the hierarchical organization of the energy landscape is
not a universal property but depends on the specific sequence and possibly on
the biological function. RNAs with high values of $u$ (e.g., F32D1.12) may
possess a rigid hierarchy promoting fast folding. RNAs with low values of $u$
(e.g., C05G6.4) have a flatter landscape, which may be associated with a need
for conformational plasticity. Under a fixed nucleotide composition, permuting
the sequence can change the degree of ultrametricity by almost an order of
magnitude (from 2.6\% to 81\%). This indicates that it is the order of
nucleotides that encodes the hierarchical structure of the landscape.
It is possible that natural sequences are not evolutionarily selected for
extreme ultrametricity but occupy an intermediate position -- presumably due
to a trade-off between folding speed and functional mobility.

We will separately discuss the possibility of generalizing the used algorithm
toward more realistic thermodynamic models. The most natural and important
direction for further research is the transition from the Nussinov model to the
more realistic Zuker thermodynamic model \cite{zuker1981,zuker2003}. The Zuker
algorithm accounts for energy contributions not only from individual pairs but
also from loops of various types and sizes, as well as neighbor-dependent
pairing energies. This allows predicting RNA secondary structures with an
accuracy substantially surpassing that of the Nussinov model. In the context
of this work, the transition to the Zuker model opens up the following
possibilities: (1) testing the stability of the obtained conclusions about
ultrametricity when moving to realistic energies; (2) investigating not only
the ground states but also sets of suboptimal structures with energy
$E_{\min}+\Delta E$; (3) analyzing the temperature dependence of
ultrametricity (from $T=0$ to melting temperatures). However, such a
transition is associated with computational difficulties: the recursions in the
Zuker algorithm account for a larger number of loop types, increasing the
computation time, and the structure space becomes sparser, which may require
increasing the sample size $M$. Nevertheless, modern implementations
(e.g., the ViennaRNA package) allow performing such calculations for RNAs up to
a few hundred nucleotides in length.

\section*{Conflict of Interest}

The author declares no conflict of interest.

\section*{Funding}

The study was carried out without external funding sources.

\section*{Data and Code Availability}

The source code of the program, as well as the sequence files in FASTA format,
are available upon request to the author.

\end{document}